\documentclass{article}
\usepackage{epsf,emulateapj,apjfonts}

\lefthead{SARAZIN \& LIEU}
\righthead{EUV FROM CLUSTERS}
\slugcomment{Astrophysical Journal Letters, in press}

\begin{document}
 
\title{Extreme Ultraviolet Emission from Clusters of Galaxies:
Inverse Compton Radiation from a Relic Population of Cosmic Ray Electrons?}
 
\author{Craig L. Sarazin}
\affil{Department of Astronomy, University of Virginia, \\
P.O. Box 3818, Charlottesville, VA 22903-0818; \\
cls7i@virginia.edu,}

\and
 
\author{Richard Lieu}
\affil{Department of Physics, University of Alabama, \\
Huntsville, AL 35899; \\
lieur@cspar.uah.edu}

\begin{abstract}
We suggest that the luminous extreme ultraviolet (EUV) emission which
has been detected recently from clusters of galaxies is Inverse
Compton (IC) scattering of Cosmic Microwave Background (CMB) radiation by
low energy cosmic ray electrons in the intracluster medium.
The cosmic ray electrons would have Lorentz factors of $\gamma \sim 300$,
and would lose energy primarily by emitting EUV radiation.
These particles have lifetimes comparable to the Hubble time;
thus, the electrons might represent a relic population of cosmic rays
produced by nonthermal activity over the history of the cluster.
The IC model naturally explains the observed increase in the ratio of
EUV to X-ray emission with radius in clusters.
The required energy in cosmic ray electrons is typically 1--10\% of the
thermal energy content of the intracluster gas.
We suggest that the cosmic ray electrons might have been produced by
supernovae in galaxies, by radio galaxies, or by particle acceleration in
intracluster shocks.
\end{abstract}


\section{INTRODUCTION} \label{sec:intro}

Recently, observations of clusters of galaxies with the
{\it Extreme Ultraviolet Explorer} ({\it EUVE}) Deep Survey instrument
using the Lexan B filter have revealed extreme ultraviolet (EUV) emission
in excess of that expected from the previously observed, X-ray emitting
intracluster medium (ICM).
The passband of this instrument is 65--248 eV.
The excess EUV emission is seen in the Virgo cluster
(Lieu et al.\ 1996b,c),
Coma cluster
(Lieu et al.\ 1996a),
Abell~1795
(Mittaz, Lieu, \& Lockman 1997),
Abell~2199,
and
Abell~4038
(Bowyer, Lieu, \& Mittaz 1997).
The observed EUV luminosities in this band, corrected for Galactic
absorption, range from
$L_{EUV} \approx 9 \times 10^{42} h_{50}^{-2}$ erg s$^{-1}$ in Virgo to
$L_{EUV} \approx 2 \times 10^{45} h_{50}^{-2}$ erg s$^{-1}$ in A1795
(where the Hubble constant is $H_o = 50 h_{50}$ km s$^{-1}$ Mpc$^{-1}$).
In all of the cases except A2199, there is an associated soft
X-ray excess seen in the {\it ROSAT} PSPC spectra of the clusters.
The ratio of {\it EUVE} to PSPC fluxes and the correction for Galactic
absorption indicate that the emitted spectrum drops rapidly with
photon energy across the EUVE band.
In A1795, the best-fit model for the spectrum implies that the average
emitted photon energy is $\langle h \nu_{EUV} \rangle \approx 75$ eV.

The clusters with detected excess EUV emission are varied in their
other properties.
Virgo is not as rich as the others.
Several have cooling flows in their centers, but Coma and A4038 do
not.
Coma has a strong radio halo, but A1795 and A2199 do not.
Virgo is the closest cluster, while A1795 is at a redshift of
$z = 0.0631$.

The EUV emission is extended over a scale comparable to the size of
the cluster.
In the three more distant clusters (A1795, A2199, and A4038),
the EUV emission falls off with radius more slowly than the
cluster X-ray emission, so that the ratio of EUV emission to
X-ray emission increases rapidly with radius.
Virgo and Coma do not show such a trend, but are closer than the
others, so the same angular scale doesn't correspond to as large
a physical radius.

In the papers reporting the discovery of this EUV emission from
clusters, a thermal emission model from diffuse gas was suggested.
The temperature of this gas would be
$T_{warm} \sim (1 - 10) \times 10^5$ K.
While it is natural to suggest such a model, given that the X-ray emission
from clusters is produced by thermal emission from diffuse gas with a
higher temperature,  there are a number of concerns with this hypothesis
(Fabian 1996).
First, a rather large mass of this warm gas is required.
In A1795, this mass is $\approx 10^{15} \, h_{50}^{-5/2} \, M_\odot$,
which is about the virial mass of the cluster
(Mittaz et al.\ 1997).
Moreover, gas at these temperatures cools very rapidly.
In A1795, the total cooling rate of gas would exceed
$10^5 \, h_{50}^{-2} \, M_\odot$ yr$^{-1}$.
There is no obvious energy source for this gas;
Mittaz et al.\ suggest accretion of intercluster gas at a very high
rate ($\ga 10^5 \, h_{50}^{-2} \, M_\odot$ yr$^{-1}$).
In this thermal model, the bulk of the radiation would be line emission.
A search for the O~VI line, which occurs at longer wavelengths but is produced
in gas of the same temperature, did not detect the line in some of the
same clusters which show EUV emission
(Dixon, Hurwitz, \& Ferguson 1996).
Finally, the observed spatial distribution of EUV emission and
the fact that the ratio of EUV to X-ray emission increases strongly with
radius is not an obvious consequence of this model.

Although a small amount of EUV emission might have been
expected in the centers of cooling flow clusters if the gas continues
to cool below X-ray emitting temperatures, this cannot be the explanation
of the observed EUV emission
(Fabian 1996).
First,
the EUV emission is much too luminous, and would imply cooling rates
$\ga$$10^2$ times higher than those seen in X-rays.
Second, not all of the clusters with EUV emission have cooling flows.
Finally, the spatial distribution of the EUV emission is the opposite
of that expected for cooling flow gas; in A1795, for example, most of
the EUV luminosity is at large radii, while the cooling flow gas is located
at the cluster center where the gas density is highest.

Fabian (1997) proposed a thermal model which avoids many of these problems.
He suggested that the warm gas was located in turbulent mixing layers
at the interface between embedded cold clouds (with $T \sim 10^4$ K)
and the ICM ($T \sim 10^8$ K).
The power source for the EUV emission is then the thermal energy content
of the ICM, which is large.
Energetically, this model works for most of the clusters, but may have
difficulties with the very luminous EUV emission of A1795.
The gas in the mixing layers is recycled, so that the cooling rates can
be greatly reduced.
It is not clear that this model would predict the extended spatial
distribution of the EUV emission when compared to the X-ray emission.

\section{INVERSE COMPTON EMISSION} \label{sec:ic}

The difficulties with the thermal model for the EUV emission have led
us to consider a nonthermal model, in which this radiation is produced
by Inverse Compton (IC) scattering of Cosmic Microwave Background (CMB)
radiation by low energy cosmic ray electrons.
Because it has the largest luminosity and posed the greatest problems for
the thermal model, we will use parameters for the EUV emission drawn from
the observations of A1795 (Mittaz et al.\ 1997).
Consider a CMB photon with
a frequency $\nu_{CMB}$ which is scattered by a relativistic electron
with an energy of $\gamma m_e c^2$, where $\gamma$ is the Lorentz factor.
On average, the resulting IC photon will have an energy of
$\langle h \nu_{IC} \rangle = (4/3) \gamma^2 \langle h \nu_{CMB} \rangle$.
The typical photon in the CMB has an energy of
$\langle h \nu_{CMB} \rangle \approx 2.8 k T_{CMB}$,
where $T_{CMB} = 2.73$ K is the temperature of the CMB.
If we associate $h \nu_{IC}$ with the average energy of EUV photons
detected from a cluster, then the average Lorentz factor of the
cosmic ray electrons is
\begin{equation} \label{eq:gamma}
\langle \gamma \rangle \approx 300
\left( \frac{\langle h \nu_{EUV} \rangle}{75 \, {\rm eV}} \right)^{1/2}
\, .
\end{equation}
Thus, the required cosmic ray electrons have very low energies compared
to the typical energies of electrons which produce radio emission in
radio sources.

There is considerable evidence that the intracluster medium (ICM) also
contains a magnetic field with a typical strength of $B \sim 1 \, \mu$G
(e.g., Rephaeli, Gruber, \& Rothschild 1987;
Kim et al.\ 1990).
The same low energy cosmic ray electrons will also generate synchrotron
radio emission with an average photon frequency of
$\langle \nu_{rad} \rangle = (55/96) (\sqrt{3}/\pi)
\gamma^2 (e B / m_e c) \sin \alpha$, where $\alpha$ is the pitch angle
of the electron.
This gives
\begin{equation} \label{eq:radio}
\langle \nu_{rad} \rangle \la
5 \times 10^5
\left( \frac{\langle \gamma \rangle}{300} \right)^2
\left( \frac{B}{1 \, \mu{\rm G}} \right) \, {\rm Hz}
\, ,
\end{equation}
where the inequality comes from the fact that $\sin \alpha \le 1$.
Thus, this radio emission would be well-below the Earth's ionospheric
cut-off, and unobservable.
The ratio of the luminosity of the synchrotron emission to that of the
IC emission is
\begin{equation} \label{eq:radio_lum}
\frac{L_{rad}}{L_{IC}} = \frac{U_{B}}{U_{CMB}}
\approx 0.095 \,
\left( \frac{B}{1 \, \mu{\rm G}} \right)^2
\, ,
\end{equation}
where $U_B$ and $U_{CMB}$ are the energy density of the magnetic field
and Cosmic Microwave Background, respectively.

This shows that synchrotron radio emission produced by the same electrons
would be unobservable. 
In particular, one would not expect that IC produced EUV
emission would be restricted to clusters with radio halos.
Such radio halos are, in fact, rather rare
(e.g., Hanisch 1982).
For example, A1795 has very strong EUV emission, but lacks a radio halo.
Models in which the EUV emission from clusters is due to the electrons
which produce the radio halos have been proposed by
Ensslin \& Biermann (1997) and
Hwang (1997).

In addition to losses by synchrotron emission, it is important to consider
if there are other energy loss processes for $\gamma \approx 300$ electrons
in the ICM which would compete with IC scattering of the CMB.
First, the ICM contains other radiation fields which might also undergo
IC scattering.
These include the near-IR---optical light from galaxies, and the X-ray
emission produced by the intracluster gas.
However, the energy densities in these radiation fields are smaller
than that in the CMB, and thus the losses are also smaller.
Low energy cosmic ray electrons are also subject to losses from
Coulomb interactions with thermal electrons in the ICM plasma
and from bremsstrahlung.
Both of these processes are proportional to the density of the
ICM plasma.
For electrons with $\gamma \approx 300$ in an ionized hydrogen plasma,
the Coulomb losses are larger than those due to bremsstrahlung.
The time scale for Coulomb energy losses is approximately
(Rephaeli 1979)
\begin{equation} \label{eq:coulomb}
t_{Coul} \approx 7 \times 10^{9} \,
\left( \frac{\gamma}{300} \right) \,
\left( \frac{n_e}{10^{-3} \, {\rm cm}^{-3}} \right)^{-1} \, {\rm yr} \, .
\end{equation}
When compared with the time scale for energy loss by IC emission
[eq.~(\ref{eq:lifetime}) below], we find that IC emission dominates
for plasma densities below a critical density $n_{crit}$,
\begin{equation} \label{eq:ncrit}
n_e \le n_{crit} \approx
10^{-3} \,
\left( \frac{\gamma}{300} \right)^2 \,
{\rm cm}^{-3} \, .
\end{equation}
This inequality is satisfied in most of the volume of a typical cluster,
and is particularly true at the outer radii where the EUV excess seems
to be strongest
(Mittaz et al.\ 1997).
However, at the centers of rich clusters, particularly in the cooling
flows, the ICM density exceeds $n_{crit}$ and Coulomb losses may be dominant.
This might help to explain the lack of detectable EUV emission in these
regions
(Mittaz et al.\ 1997),
although the brightnesses of the soft X-ray emission there also makes
it more difficult to separate EUV and X-ray emission.

IC losses will dominate for the low energy cosmic rays needed to
produce the observed EUV emission from clusters.
The lifetime of these particles is then
\begin{equation} \label{eq:lifetime}
t_{IC} = \frac{\gamma m_e c^2}{\frac{4}{3} \sigma_T c \gamma^2 U_{CMB}} =
7.7 \times 10^{9}
\left( \frac{\gamma}{300} \right)^{-1} \,
{\rm yr}
\, .
\end{equation}
Thus, the lifetime of a $\gamma \approx 300$ electron is approximately
the Hubble time, or the likely average age of clusters.
We would then expect that any low energy cosmic rays with
$\gamma \approx 300$ which were generated within or had escaped to
the ICM over the lifetime of a cluster would still be present there.

Note that equations~(\ref{eq:gamma}) and (\ref{eq:lifetime}) depend only
on the properties of the CMB.
Thus, these arguments show that the EUV spectral band is a unique region
in which one can detect nonthermal emission from an accumulated relic
population of cosmic rays due to the total history of particle acceleration
and activity over most of the lifetime of the universe in any diffuse
regions (with $n_e \la n_{crit}$).
To our knowledge, this unique role of the EUV spectral band has not been
appreciated previously.
On the other hand, the energy density of the CMB increases with redshift
as $U_{CMB} \propto ( 1 + z )^4$.
The lifetime for IC losses becomes much shorter than the Hubble time
for redshifts $ z \la 2$.

\section{REQUIRED ENERGY IN COSMIC RAYS} \label{sec:energy}

Let $n_{CR} ( \gamma ) d \gamma$ be the number density of cosmic ray
electrons with Lorentz factors between $\gamma$ and $\gamma + d \gamma$.
Define an average Lorentz factor as
\begin{equation} \label{eq:average_gamma}
\langle \gamma \rangle \equiv
\frac{\int n_{cr}(\gamma) \gamma^2 \, d \gamma}
{\int n_{cr}(\gamma) \gamma \, d \gamma}
\, .
\end{equation}
Then, the IC luminosity of the cluster is
\begin{equation} \label{eq:luminosity}
L_{IC} = \frac{4}{3} \,
\frac{\sigma_T}{m_e c} \, \langle \gamma \rangle U_{CMB} E_{CR}
\, ,
\end{equation}
where $E_{CR}$ is the total energy of the cosmic ray electrons in the
cluster.
If the Inverse Compton luminosity is equated to the observed EUV emission,
the required total energy in cosmic rays electrons is found to be
\begin{equation} \label{eq:energy_cr}
E_{CR} \approx 2.4 \times 10^{62}
\left( \frac{L_{EUV}}{10^{45} \, {\rm erg} \, {\rm s}^{-1}} \right)
\left( \frac{ \langle \gamma \rangle }{300} \right)^{-1} \, {\rm erg} \, .
\end{equation}

It is useful to compare the energy in cosmic rays to that in the thermal
energy $E_{gas}$ of the intracluster gas.
This ratio is
\begin{eqnarray}
\frac{E_{CR}}{E_{gas}} & = & 0.085
\left( \frac{L_{EUV}}{10^{45} \, {\rm erg} \, {\rm s}^{-1}} \right)
\left( \frac{ \langle \gamma \rangle }{300} \right)^{-1} \nonumber \\
& & \qquad \times \left( \frac{M_{gas}}{10^{14} \, M_\odot} \right)^{-1}
\left( \frac{T}{7 \times 10^7 \, {\rm K}} \right)^{-1}
\, , \label{eq:energy_ratio}
\end{eqnarray}
where $M_{gas}$ is the mass of the ICM, and $T$ is its average temperature.
The ratio of pressures is lower by a factor of two, since the cosmic
rays are relativistic and the thermal gas is not.
Thus, the production of the observed EUV emission in clusters requires
a relic cosmic ray electron population with an energy density and pressure which
are 1--10\% of that of the thermal gas.

One concern is that the inclusion of cosmic ray ions may increase the total
CR energy significantly.
If the cosmic rays have a very steep energy spectrum, and if the number
of cosmic ray ions at 150 MeV ($\gamma \approx 300$ in electrons) is
reduced because these particles are not relativistic, then the increase
in pressure due to ions might only be a factor of a few.
On the other hand, cosmic ray ions outnumber electrons by a factor of
$\sim$10$^2$ at the top of the Earth's atmosphere
(e.g., Webber 1983).

\section{SPATIAL DISTRIBUTION OF EUV EMISSION} \label{sec:spatial}

The Inverse Compton model naturally explains the spatial distribution of
the EUV emission and the fact the EUV to X-ray ratio increases rapidly
with radius.
The X-ray emission from the ICM is due to electron-ion collisions,
and its emissivity is proportional to the square of the gas density.
The Inverse Compton emission is due to collisions between cosmic
ray electrons and CMB photons;
the distribution of the latter is uniform.
Thus, the Inverse Compton emissivity is linearly proportional to density,
and the resulting surface brightness will decline with radius less
rapidly than that for the X-ray emission.

As a specific but probably oversimplified model, let us assume that
the pressure of the cosmic ray electrons, $P_{CR}$, is proportional
to the thermal gas pressure $P_{gas}$.
Further assume that the energy spectrum of the cosmic rays is independent
of radius.
For this calculation, we will assume that the thermal gas pressure in a
cluster is represented by the ``beta model'' with
$P_{gas} (r) \propto [ 1 + ( r / r_c ) ]^{-3 \beta /2}$.
Here, $r$ is the radius and $r_c$ is the core radius.
Then, the EUV surface brightness varies as
$I_{EUV} (r) \propto [ 1 + ( r / r_c ) ]^{-3 \beta /2 + 1/2}$, while
the X-ray surface brightness varies as
$I_{X} (r) \propto [ 1 + ( r / r_c ) ]^{-3 \beta + 1/2}$.
The ratio of the EUV to X-ray
surface brightness goes as
$I_{EUV}/I_{X} \propto 1/P_{gas} \propto [ 1 + ( r / r_c ) ]^{3 \beta /2}$.

\centerline{\null}
\vskip2.65truein
\includegraphics{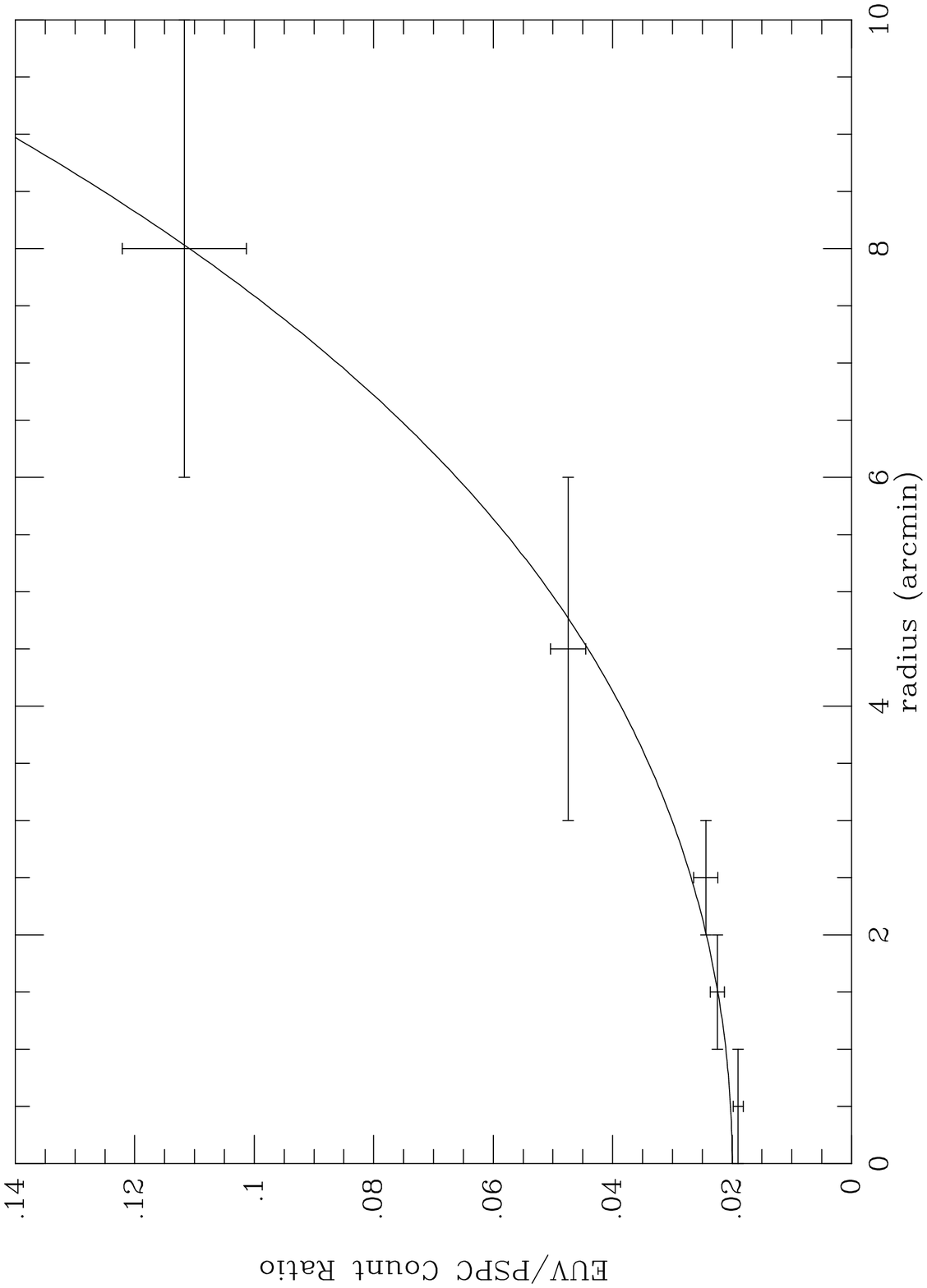}
\figcaption{The ratio of EUV to X-ray surface brightness in A1795 as a function of
projected radius (Mittaz et al.\ 1997).
The solid curve is the ``beta model'' prediction for this ratio in the
Inverse Compton theory, assuming a constant cosmic ray to gas pressure
ratio.
The parameters of the beta model are taken from the fit to the
X-ray surface brightness by Briel \& Henry (1996).
\label{fig:euv_xray}}

\vskip0.2truein

Figure~\ref{fig:euv_xray} shows the observed ratio of EUV to X-ray
surface brightness measured for several annuli in A1795
(Mittaz et al.\ 1997).
For comparison, the results of this simple beta model are shown.
The parameters of the beta model ($r_c$ and $\beta$) were taken
from the fit to the observed X-ray emission by
Briel \& Henry (1996);
there are no free parameters in this model except for the overall
normalization.
Given the many simplifying assumptions in this model, the detailed agreement
is probably fortuitous.
However, the comparison does show that the Inverse Compton model can explain
the qualitative radial behavior of the EUV surface brightness.

\section{DISCUSSION} \label{sec:discussion}

Normally, nonthermal emission processes produce broad power-law spectra,
partly as a result of power-law energy spectrum of the cosmic ray
electrons.
The observations of clusters suggest that the EUV emission drops
rapidly in going from EUV to soft X-ray photon energies.
On the low energy side, the total energies involved would be large and
the surface brightnesses would violate observational limits if the
spectrum continued to rise down to the near UV and optical bands.
However, there are reasons to think that the CR electron population
in clusters might peak near the energies required to produce the
EUV emission ($\gamma \approx 300$).
On the higher energy side, the electron population will be depleted by
IC losses over the age of the cluster
[eq.~(\ref{eq:lifetime})].
For continual particle injection, this steepens the exponent of a power-law
energy spectrum by unity.
If most of the CRs were injected early in the history of the cluster,
the effect would be larger.
On the lower energy side of the CR spectrum, electrons will lose energy
rapidly to Coulomb losses even at low ICM densities
[eq.~(\ref{eq:coulomb})].

What is the ultimate source of these relic cosmic ray electrons, which have
accumulated over the lifetime of the cluster?
Note that we typically need $E_{CR} \sim 3 \times 10^{61}$ erg, although the
values in A1795 are a factor of 10 larger
[eq.~(\ref{eq:energy_cr})].
This implies that the ratio of the cosmic ray electron energy to that
of the thermal gas is typically $\sim$1\%, but with a value of
$\sim$10\% in A1795.

One possible source for these cosmic rays would be supernovae, supernova
remnants, and pulsars associated with the stellar population in the
galaxies.
The present rate of star formation in the early-type galaxies in rich
clusters is rather low, and the Type Ia supernovae produced by the
older stellar population occur at too low a rate to give the required
number of cosmic rays.
However, it is believed that these galaxies had much higher supernova
rates in the past.
The heavy element abundances in clusters provide a useful constraint on
the overall number of such supernovae.
As a simple example, let us assume that each supernova produces 0.5
$M_\odot$ of heavy elements and a total cosmic ray electron energy
(from the supernova, its remnant, pulsar, etc.) of $10^{49}$ erg.
Then, the production of a half-solar heavy element abundance in the
ICM will generate a cosmic ray electron population of
$E_{CR}/E_{gas} \sim$ 1\%.
This might be barely adequate.
One concern is that the cosmic ray electrons might experience strong
Coulomb or adiabatic losses before they reach the ICM.
Another worry is that the early-type galaxies in clusters may have
experienced most of their star formation and supernovae before $z \sim 2$,
and the cosmic rays would have also undergone strong IC losses.
A prediction of this model for the origin is that the EUV properties of
clusters might be relatively uniform from cluster to cluster, just as the
abundances of heavy elements in the ICM are fairly constant.

Another possibility is that the cosmic rays may have come from radio galaxies,
quasars, or other AGNs in the cluster over its lifetime.
The total cosmic ray electron energy required corresponds to 10--100
Cyg-A radio sources, and the present day radio luminosity function of
clusters suggests that there are too few bright radio galaxies to
produce the cosmic rays.
However, the energy content of radio sources and their radio luminosities
are not necessarily well-correlated.
Moreover, there is evidence that both the general population of luminous
radio galaxies and quasars was higher in the past, and that they were
more commonly associated with clusters of galaxies.
In this model, one might expect large variations in the EUV luminosity
of otherwise similar clusters, which would be associated with the small
number statistics of particularly energetic AGN.

Finally, the cosmic ray electrons might have been generated {\it in situ}
in the ICM by particle acceleration associated with shocks and/or strong
turbulence.
Unless the ICM underwent strong preheating, most of the gas must have
passed through high Mach number shocks with velocities $\sim 10^3$ km
s$^{-1}$.
These would include overall cluster accretion shocks (if much of the
cluster formed from a single large perturbation) and/or subcluster merger
shocks if the cluster formed hierarchically.
In the interstellar medium of our Galaxy, shocks of this velocity
associated with supernova remnants also produce cosmic ray electrons with
about 1--10\% of the shock energy.
This could lead to a population of such electrons in the ICM with
$E_{CR}/E_{gas} \sim$ 1--10\%.
In this model, modest variations in the EUV luminosity from cluster to
cluster might be expected as a result of variations in their detailed
histories.

The IC model for the EUV emission of clusters predicts a continuum spectrum
for the emission.
On the other hand, in the thermal model most of the emission is in lines.
Thus, spectra of this component could decide between thermal and nonthermal
models.
The cosmic ray electrons required by the IC model would also generate some
gamma-ray emission via the scattering of higher energy photons and
bremsstrahlung.
The expected fluxes are $10^{-12} - 10^{-11}$ erg cm$^{-2}$ s$^{-1}$
at 0.3 MeV (from IC scattering of optical photons) and at 150 MeV
(from bremsstrahlung).
These predicted fluxes are near the sensitivity limits with current
instrumentation, but should be detectable in the future.

\acknowledgements

We thank Ian Axford, Andy Fabian, and an anonymous referee
for useful comments.
R. L. thanks Torsten Ensslin and Chorng-yuan Hwang for helpful discussions.
C. L. S. was supported in part by
NASA ROSAT grants NAG 5--3308 and NAG 5--4787,
NASA ASCA grants NAG 5--2526 and NAG 5--4516,
and NASA Astrophysical Theory Program grant 5-3057.
R. L. was supported by NASA EUVE and ADP grants 5-34374 and 5-34378 awarded
to the University of Alabama, Huntsville.

\end{document}